\documentclass[reprint,amsmath,amssymb,aps,prb,]{revtex4-2}

\usepackage{graphicx}
\usepackage{amsmath}
\usepackage{xcolor}
\usepackage{dcolumn}
\usepackage{enumitem}
\usepackage{epstopdf}

\begin{document}

\title{Exchange spin waves in thin films with gradient composition}

\author{I.A. Golovchanskiy$^{1,2,3,*}$,  I.V. Yanilkin$^{4,5}$, A.I. Gumarov$^{4,5}$, B.F. Gabbasov$^{4,5}$, N.N. Abramov$^{2}$, R.V. Yusupov$^{4}$, R.I. Khaibullin$^{5}$, V.S. Stolyarov$^{1,2,3}$, L.R. Tagirov$^{4,5,**}$}

\affiliation{
$^1$ Moscow Institute of Physics and Technology, State University, 9 Institutskiy per., Dolgoprudny, Moscow Region 141700, Russia; 
$^2$ National University of Science and Technology MISIS, 4 Leninsky prosp., Moscow 119049, Russia; 
$^3$ Dukhov Research Institute of Automatics (VNIIA), 127055 Moscow, Russia; 
$^4$ Institute of Physics, Kazan Federal University, Kremlyovskaya str. 18, 420008 Kazan, Russia; 
$^5$ Zavoisky Physical-Technical Institute, FRC Kazan Scientific Centre of RAS, 420029 Kazan, Russia;\\
}%

\begin{abstract}
We report investigation of ferromagnetic resonance phenomenon in ferromagnetic thin films with essentially non-uniform composition.
Epitaxial Pd-Fe thin film with linear distribution of Fe content across the thickness is used as the model material.
Anomalous perpendicular standing spin waves are observed and quantified using the collective dynamic equation.  
Numerical analysis yields the exchange stiffness constant for diluted Pd-Fe alloy $D=2A/\mu_0M_s=15$~T$\cdot$nm$^2$ and the ratio of the effective magnetization to the saturation magnetization $M_{eff}/M_s=1.16$.
It is demonstrated that, overall, engineering of thin films with non-uniform composition across the thickness can be used for high-frequency or low-field magnonic operations using exchange spin waves.
\end{abstract}

\maketitle

\section{Introduction}

Magnonics is an expanding field of research that offers approaches for transmission and processing of microwave signals via spin waves (i.e., magnons) \cite{Kruglyak_JPDAP_43_264001, Serga_JPDAP_43_264002, Lenk_PhysRep_507_107, Chumak_NatPhys_11_453, Csaba_PLA_381_1471, Barman_JPCM_33_413001, Chumak_arXiv_2111_00365, Pirro_NatRevMat_6_1114}. 
Practical advantages of the magnonics include tunability of the magnon frequencies via the external field, material choice, or the geometry of the magnon media, charge-current-free nature and low power consumption of the spin-wave transfer, and micro- and sub-micro-scale dimensions of spin waves at microwave frequencies, which allows to target creation of micro-devices for processing of microwave information.
Currently, the applied magnonics is progressed towards development of magnon logic devices\cite{Chumak_arXiv_2111_00365}, i.g., waveguides \cite{Wessels_SciRep_6_22117}, magnon transistors \cite{Chumak_NatComm_5_4700}, directional couplers \cite{Wang_NatEl_3_765}, majority gates \cite{Klingler_APL_106_212406,Ganzhorn_APL_109_022405}, non-Boolean devices \cite{Csaba_PLA_381_1471}, and neuromorphic circuits \cite{Sadovnikov_JETPLett_108_312,Papp_NatComm_12_6422}.
Also, magnonics finds its application in spintronic systems\cite{Saitoh_APL_88_182509,Kajiwara_Nature_464_262,Zhang_JAP_117_17C727,Chumak_NatPhys_11_453,Vaidya_Sci_368_160}.
Precessing magnetization is used as the source of the spin current in a conductor that is induced by the spin pumping at the interface of the ferromagnet/conductor bilayer.
The charge current is subsequently converted from the spin current via the inverse spin-Hall effect in adjacent conductor. 
As the alternative direction of research, cavity magnonics \cite{Huebl_PRL_111_127003,Tabuchi_PRL_113_083603,Zhang_PRL_113_156401,Hou_PRL_123_107702,Li_PRL_123_107701,Golovchanskiy_SciAdv_7_eabe8638,Golovchanskiy_PRAppl_16_034029} is progressed towards operation with the single quantum and considers hybridization of magnons with photons.
Cavity magnonics offers various promising technologies, including hybrid quantum platforms \cite{Tabuchi_Sci_349_405,Lachance-Quirion_Sci_367_425}, magnon memory \cite{Zhang_NatComm_6_8914}, and microwave-to-optical quantum transducers \cite{Hisatomi_PRB_93_174427}.

Arguably, one of the most desired characteristics regardless of particular application is high magnon frequency at zero or low magnetic fields.
Quite a number of approaches have been proposed in recent years that allow to achieve high frequencies at low fields.
For instance, the enhanced eigen frequencies of a thin-film magnonic media can be accessed by engineering the magnetic anisotropy at interfaces of multilayered thin films \cite{Johnson_RPP_59_1409,Haldar_SciAdv_3_e1700638,Golovchanskiy_PRAppl_11_044076,Lau_PRM_3_104419}, by considering antiferromagnetic exchange interactions \cite{Rezende_JAP_126_151101,MacNeill_PRL_123_047204,Golovchanskiy_arXiv_2108_03847,Vaidya_Sci_368_160}, or using mechanisms of hybridization with superconducting structures \cite{Li_ChPL_35_077401,Golovchanskiy_PRAppl_14_024086,Golovchanskiy_AdvFuncMater_28_1802375}.

Technically, high eigen-frequencies in the magnonic media at low magnetic fields can be reached much easier by exploring exchange spin waves.
Indeed, for perpendicular standing spin waves in ferromagnetic films \cite{Kittel_PR_100_1295,Seavey_JAP_30_S227,Klingler_JPDAP_48_015001} the effective exchange field $H_e$ scales with the thickness $d$ as $H_{ex}=(2A/\mu_0M_s)k^2$, where $A$ is the exchange stiffness constant, $M_s$ is the saturation magnetization, $k=2\pi n/d$ is the wavevector, and $n$ is the mode number.
Appropriate choice of ferromagnetic material or of just the film thickness allows to achieve eigen-frequencies in the range from GHz up to sub-THz.
However, conventionally, coupling of perpendicular exchange spin waves in thin films with microwave fields requires closed boundary conditions for spin waves at least at one of film surfaces.
Otherwise, the dynamics susceptibility of the perpendicular spin waves is zero.

Often spin-wave boundary conditions in ferromagnetic films are opened and the operation with exchange spin waves requires more sophisticated approached.
For instance, as demonstrated recently, excitation of exchange spin waves can be achieved using the dipole or the interface exchange interactions in magnon-magnon hybrid structures \cite{Yu_NatComm_7_11255,Klingler_PRL_120_127201,Chen_PRL_120_217202,Liu_NatComm_9_738,Li_PRL_124_117202} or by employing spin transfer torque nano-oscillators \cite{Demidov_NatMat_9_984,Madami_NatNano_6_635,Urazhdin_NatNano_9_509}.
More exotic approaches for excitation of exchange spin waves include employment of moving superconducting vortex lattice \cite{Dobrovolskiy_arXiv_2103_10156} or resonating skyrmions \cite{Chen_ACSNano_15_4372}.

In this work, we propose and explore a more traditional approach for excitation of exchange spin waves and tuning of the effective exchange energy.
We study exchange spin waves in thin films with non-uniform composition across the thickness.
The non-uniform composition ensures the non-zero dynamic susceptibility of perpenducular spin waves regardless the boundary conditions, while particular profile of the non-uniform composition determines the effective field.
With modern development of thin film deposition techniques, opportunities for fabrication of thin films with predefined distribution of ferromagnetic parameters across the thickness are vast.

We perform our proof-of-principle measurements on epitaxially-grown Pd-Fe thin film alloy as the model material.
Diluted Pd-Fe alloys belong to a class of ferromagnets composed of palladium or platinum doped by transition metals (i.e., Fe, Co, and Ni).
Low concentration of transition metals induces spontaneous magnetization of alloys with mK Curie temperature and with a ``giant'' effective magnetic moment per magnetic atom \cite{Buscher_PRB_46_983}.
Such diluted allows are of a particular interest in superconducting spintronics \cite{Bolginov_JETPLett_95_366,Vernik_IEEE_23_1701208,Karelina_JETPLett_112_705,Karelina_JAP_130_173901,Golovchanskiy_JAP_120_163902,Esmaeili_SciChiMat_64_1246,Mohammed_Beil_11_807,Yanilkin_Nanomat_11_64}.
For higher concentrations, the magnetization and the Curie temperature increase approximately linearly with increasing Fe concentration up to 400~K for 20 at.\% of Fe \cite{Ewerlin_JPCM_25_266001,Esmaeili_SciChiMat_64_1246}.
At last, strong ferromagnetic super-structured Fe-Pd, Fe-Pt, Co-Pt films with equiatomic composition are characterized by large magneto-crystalline anisotropies\cite{Ristau_JAP_86_4527,Golovchanskiy_JPDAP_46_215502} and find its application in hard-drives \cite{Kryder_IEEE_96_1811}, micro-devices \cite{Bader_RMP_78_1,Ho_JAP_117_213901}, and high-frequency ferromagnetic resonance applications \cite{Becker_APL_104_152412, Golovchanskiy_SUST_30_054005}.
Spin-wave dynamics studied in this work contributes to understanding of magnetic phenomena in Pd-Fe system, as well as enriches the range of potential applications of Pd-Fe alloys.

\section{Experimental details}

The epitaxial Pd-Fe film samples with gradient composition across their thickness were fabricated by molecular-beam epitaxy (MBE) in ultrahigh vacuum conditions of $5\times10^{-10}$~mbar.
Epi-ready ($R_a<0.5$ nm) (100)-MgO single-crystal substrates were used with dimensions of $5\times10\times0.5$ mm$^3$ provided by \emph{Crystal GmbH}. 
Iron and palladium were evaporated simultaneously from \emph{Createc} effusion cells with precise temperature control of $\pm0.1^\circ$C. 
During the deposition process, the temperature of the palladium was constant, while the temperature of the iron was varied over time to obtain a profile of Fe content across the thickness of synthesized films. 
Film growth was carried out in two stages, by analogy with Ref. \cite{Esmaeili_TSF_669_338}. 
At the first stage, the substrate temperature was 400$^\circ$C and the first layer of the Pd-Fe alloy of 3 nm thick was deposited. 
At the second stage, the substrate temperature was 150$^\circ$C and a final layer of 113 nm thick was deposited. 
The structural perfection of the film was monitored \emph{in-situ} at each stage by low-energy electron diffraction. 

Next, the synthesized samples were cut into several pieces for further studies.
One piece of each sample was additionally annealed in ultrahigh vacuum ($9\times10^{-9}$~mbar) at a temperature of 600$^\circ$C for 2 hours in order to confirm stability of samples:  
structural and magnetic properties of samples did not change noticeably after the annealing. 

The crystal structure was studied using X-ray Bruker D8 Advance diffractometer. 
XRD studies confirmed the epitaxial growth of a Pd-Fe film alloy on MgO single-crystal substrate as ``cube-on-cube''. 
More details on structure and technology of studied films are provided in supplementary.

\begin{figure}[!ht]
\begin{center}
\includegraphics[width=0.99\columnwidth]{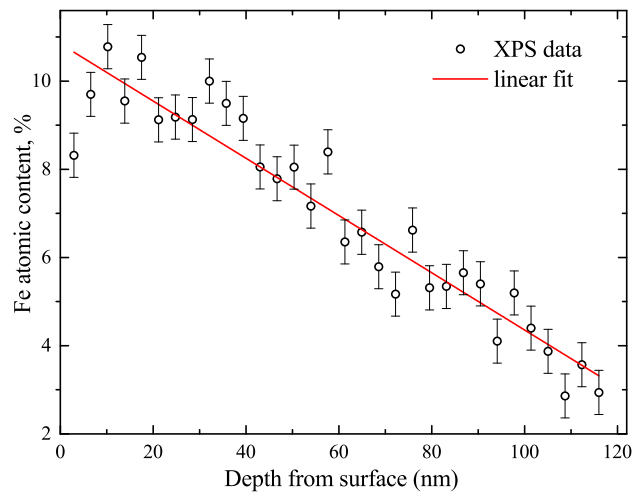}
\caption{The dependence of Fe at.\% content across the thickness of studied Pd-Fe film.
Red line shows the linear fit. }
\label{XPS}
\end{center}
\end{figure}

As a result, a series of thin film samples was fabricated of various thicknesses and compositions where the content of Fe is varied linearly across the thickness.
In this work, as the representative example, we focus on the Pd-Fe sample of thickness 116~nm where the concentration of Fe is varied linearly from 3.3~at.\% at the interface with the substrate up to 10.8~at.\% at the free surface.
The distribution of Fe atomic content across the thickness is verified with X-ray photoelectron spectroscopy in combination with sequential Ar-ion etching of the sample surface and is shown in Fig.~\ref{XPS}.

Magnetization measurements were performed using vibrating-sample magnetometer (VSM) on a \emph{Quantum Design PPMS-9} setup. 
The external magnetic field was applied both parallel to the film plane (in-plane) along [100] and [110] crystallographic directions of the cubic MgO single-crystal, and perpendicular (out-of-plane) to the film plane, along [001] direction.
Magnetization measurements reveal the Curie temperature about $T_c=250$~K for selected Pd-Fe sample.

Ferromagnetic resonance spectroscopy (refereed to as FMR) was performed using the VNA-FMR flip-chip approach \cite{Neudecker_JMMM_307_148,Kalarickal_JAP_99_093909}.
Pd-Fe sample was glued on top of the transmission line of the 50-Ohm-impedance coplanar waveguide equipped with SMP rf connectors and is placed in a home-made superconducting solenoid inside a closed-cycle cryostat (Oxford Instruments Triton, base temperature 1.2 K). 
Magnetic field is applied out-of-plane, perpendicular to the surface of the waveguide and of the studied Pd-Fe sample (i.e., along [001] direction). 
The response of experimental samples is studied by analyzing the transmitted microwave signal $S_{21}(f,H)$ with the VNA Rohde \& Schwarz ZVB20.

Cavity magnetic resonance (CMR) spectroscopy was performed using commercial X-band \emph{Bruker ESP300} electron spin resonance spectrometer equipped with the \emph{Oxford Instruments ESR-9} helium flow cryostat. 
The standard ER4102ST rectangular TE102-mode cavity was used in the measurements. 
See Ref. \cite{Esmaeili_AppMagRes_49_175} for more details on the measurement setup and its application for studies of Pd-Fe films.

\section{Experimental results and discussion}

\subsection{Magnetic properties of Pd-Fe epitaxial films with gradient composition}

\begin{figure}[!ht]
\begin{center}
\includegraphics[width=0.99\columnwidth]{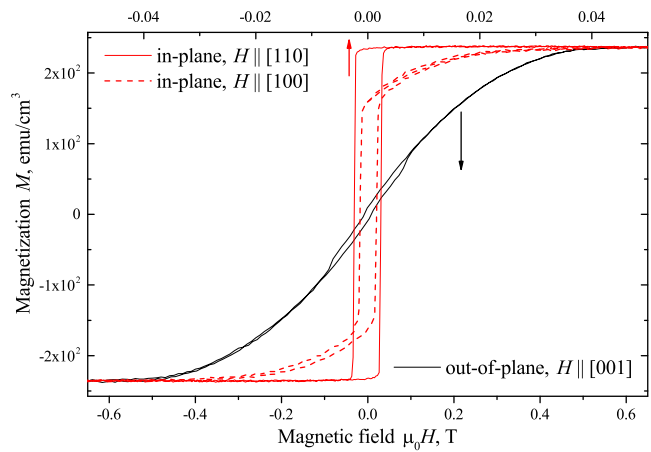}
\caption{The dependence of magnetization of Pd-Fe sample on the magnetic field in both the in-plane (red curves, upper $x$-scale) and the out-of-plane (black curve, lower $x$-scale) measurement geometries.
}
\label{M_Vs_H}
\end{center}
\end{figure}

Figure~\ref{M_Vs_H} shows the dependence of magnetization of the Pd-Fe sample on magnetic field at $T=5$~K for both the in-plane (red curve) and the out-of-plane (black curve) orientations of the magnetic field.
All curves indicate a minor hysteresis with the coercive field of about 1~mT, and the volume-averaged saturation magnetization of the Pd-Fe sample about $\overline{M}_s=236$~emu/cm$^3$ or $\mu_0\overline{M}_s=0.3$~T.
The difference in saturation fields for in-plane magnetization curves indicates the [110] crystal axis as the easy magneto-crystalline cubic axis.
The saturation field for the out-of-plane geometry $H_s$ corresponds to some overall effective magnetization $\mu_0H_s=\mu_0\overline{M}_{eff}\approx 0.47$~T, which incorporates the saturation magnetization as well as averaged uniaxial and cubic anisotropies \cite{Golovchanskiy_PRAppl_11_044076,Klingler_JPDAP_48_015001}.

\begin{figure}[!ht]
\begin{center}
\includegraphics[width=0.99\columnwidth]{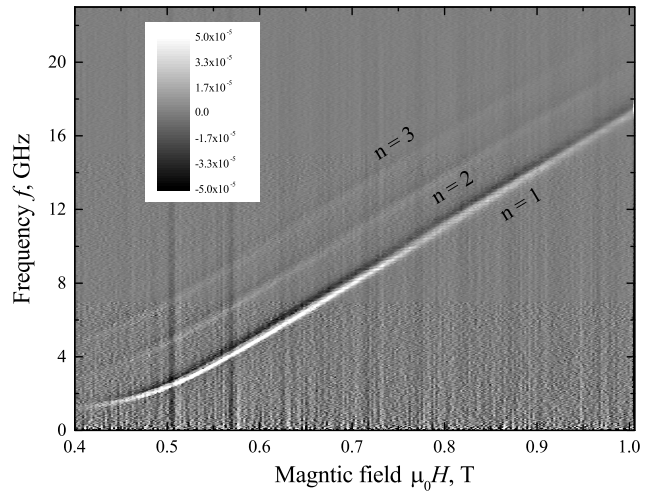}
\caption{Differentiated FMR absorption spectrum $dS_{21}(f,H)/dH$ of the Pd-Fe sample at $T=5$~K.}
\label{FMR_exp}
\end{center}
\end{figure}

Figure~\ref{FMR_exp} shows the differentiated FMR spectrum $dS_{21}/dH$ of the same PdFe sample measured at out-of-plane magnetic field and at $T=5$~K.
The spectrum reveals three absorption lines indicated with numbers $n=1,2,3$.
The line $n=1$ shows the highest absorption amplitude, though, low FMR signal does not allow to estimate relative intensities of spectral lines.
The dependence of the FMR frequency on the magnetic field shows linear dependence at $\mu_0H\gtrsim 0.5$~T following typical FMR relation for saturated thin films at out-of-plane magnetic fields.

\begin{figure}[!ht]
\begin{center}
\includegraphics[width=0.99\columnwidth]{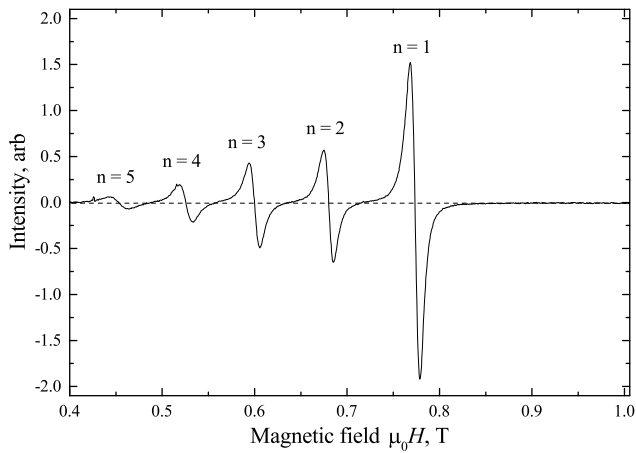}
\caption{CMR spectrum of the Pd-Fe sample at $T=5$~K.}
\label{EPR_exp}
\end{center}
\end{figure}

Relative intensities of resonance lines are obtained by measuring the cavity resonance spectrum.
Figure~\ref{EPR_exp} shows the CMR spectrum of the same Pd-Fe sample at out-of-plane magnetic field at temperature $T=5$~K and cavity frequency 9.44~GHz.
The spectrum reveals five resonance fields indicated with numbers $n=1,..,5$.
Intensities of resonances gradually reduce with increasing the mode number.


%
\begin{figure}[!ht]
\begin{center}
\includegraphics[width=0.99\columnwidth]{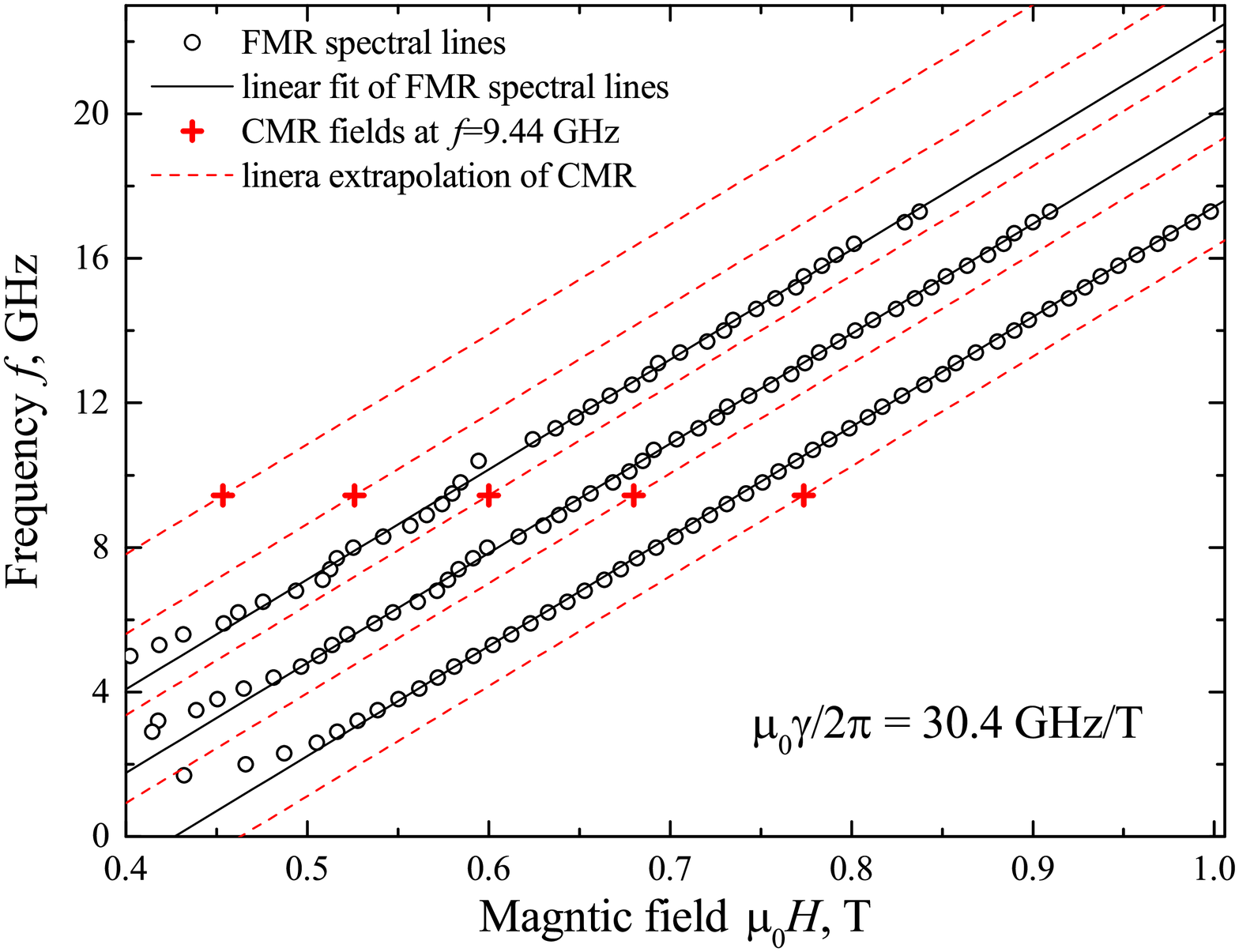}
\caption{Experimental $f_r(H)$ dependencies acquired from the FMR spectrum in Fig.~\ref{FMR_exp} are shown with black circles.
Black solid lines show the fit of FMR $f_r(H)$ lines with Eq.~\ref{kit}.
Resonance fields obtained by the CMR measurements at $f_r=9.44$~GHz are shown with red crosses.
Red dashed lines show extrapolation of CMR using Eq.~\ref{kit} and the gyromagnetic ratio obtained from the FMR lines.}
\label{FMR_sum}
\end{center}
\end{figure}

Figure~\ref{FMR_sum} summarises ferromagnetic resonance measurements.
Symbols in Fig.~\ref{FMR_sum} show experimental FMR and CMR $f_r(H)$ dependencies.
In general, FMR in thin films at out-of-plane magnetic field above the saturation field obey the modified Kittel formula \cite{Kittel_PR_100_1295,Seavey_JAP_30_S227,Klingler_JPDAP_48_015001}.
\begin{equation}
2\pi f_r/\mu_0\gamma=H+H_{ex}-M_{eff},
\label{kit}
\end{equation}
where $\gamma$ is the gyromagnetic ratio of Fe-Pt and $H_{ex}$ is the effective exchange field. 
Fitting FMR lines at $\mu_0H>0.5$~T yields the gyromagnetic ratio $\mu_0\gamma/2\pi=30.4$~GHz/T and the effective field term $\mu_0(M_{eff}-H_{ex})=0.43$~T for $n=1$, while CMR $n=1$ resonance field yields $\mu_0(M_{eff}-H_{ex})=0.46$~T.
The gyromagnetic ratio of Pd-Fe sample is close to the ratio for free electrons 28~GHz/T. 
Its slightly higher value can be explained within the two sub-lattice model\cite{Wangsness_PRB_91_1085,Hardison_JdPC_32_C1} by larger g-factor for Pd than for Fe.

The obvious shift between CMR and FMR resonances in Fig.~\ref{FMR_sum} by $\mu_0H=0.03$~T indicates presence of additional exchange field in case of FMR measurements. 
We state that this exchange field appears due to the ferromagnetic skin effect \cite{Kostylev_JAP_115_173903,Lin_JAP_117_053908,Flovik_JAP_119_163903}.
In certain conditions, in case of asymmetric excitation of magnetization dynamics, as done by the coplanar waveguide in FMR measurements, microwave currents induced in conducting ferromagnetic films partially screen external microwave fields.
This localized screening enables the excitation of magnetization dynamics with non-zero wave vector, which contributes into the process by additional exchange field and shifts the resonance frequency as compared to the Kittel mode.
In case of the CMR measurements excitation microwave fields are applied symmetrically at both surfaces of the film and the localized screening process is not enabled.
Below we focus on CMR data for simplicity.

\subsection{Exchange spin waves in Pd-Fe epitaxial films with gradient composition}

\begin{figure*}[!ht]
\begin{center}
\includegraphics[width=0.67\columnwidth]{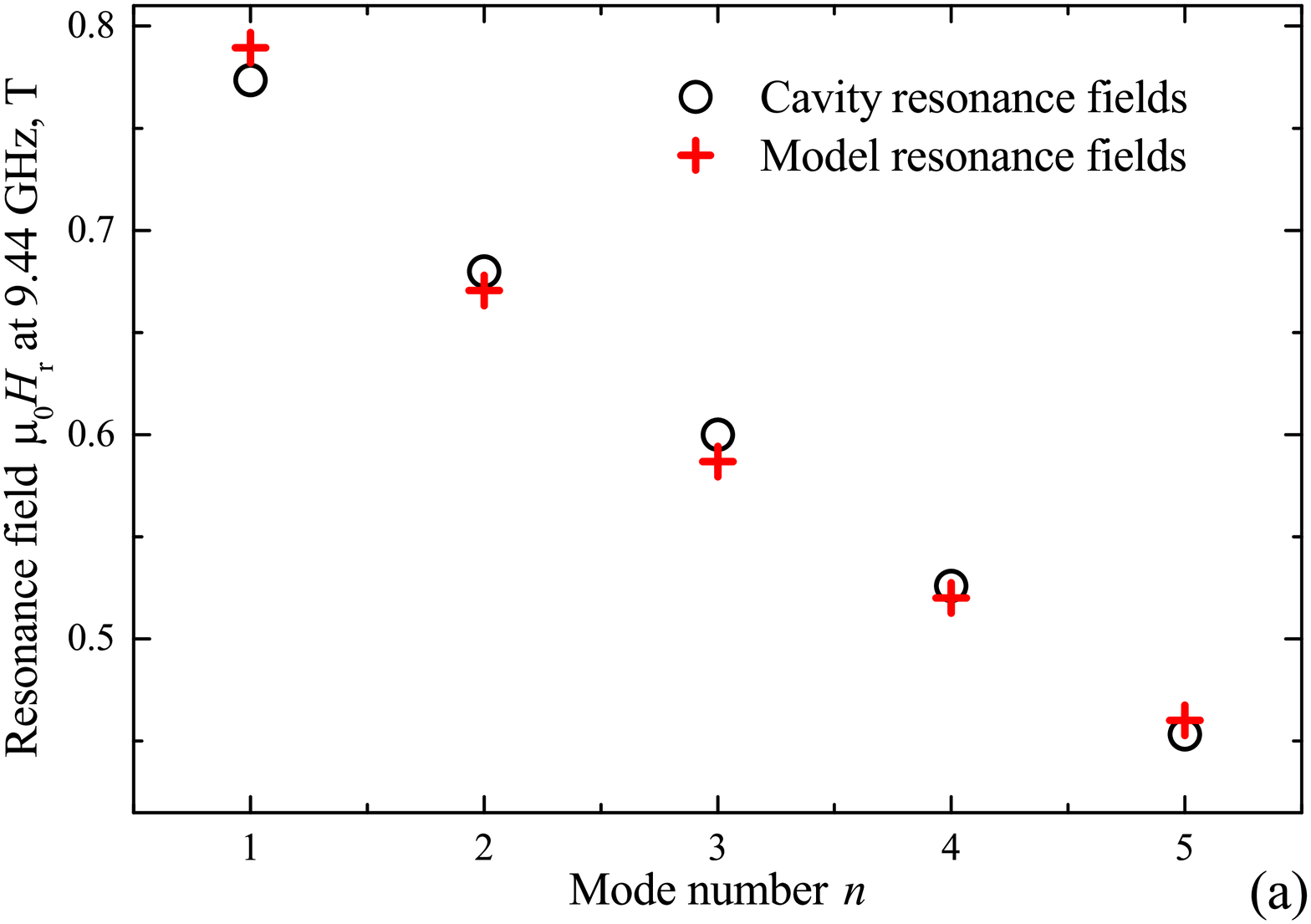}
\includegraphics[width=0.67\columnwidth]{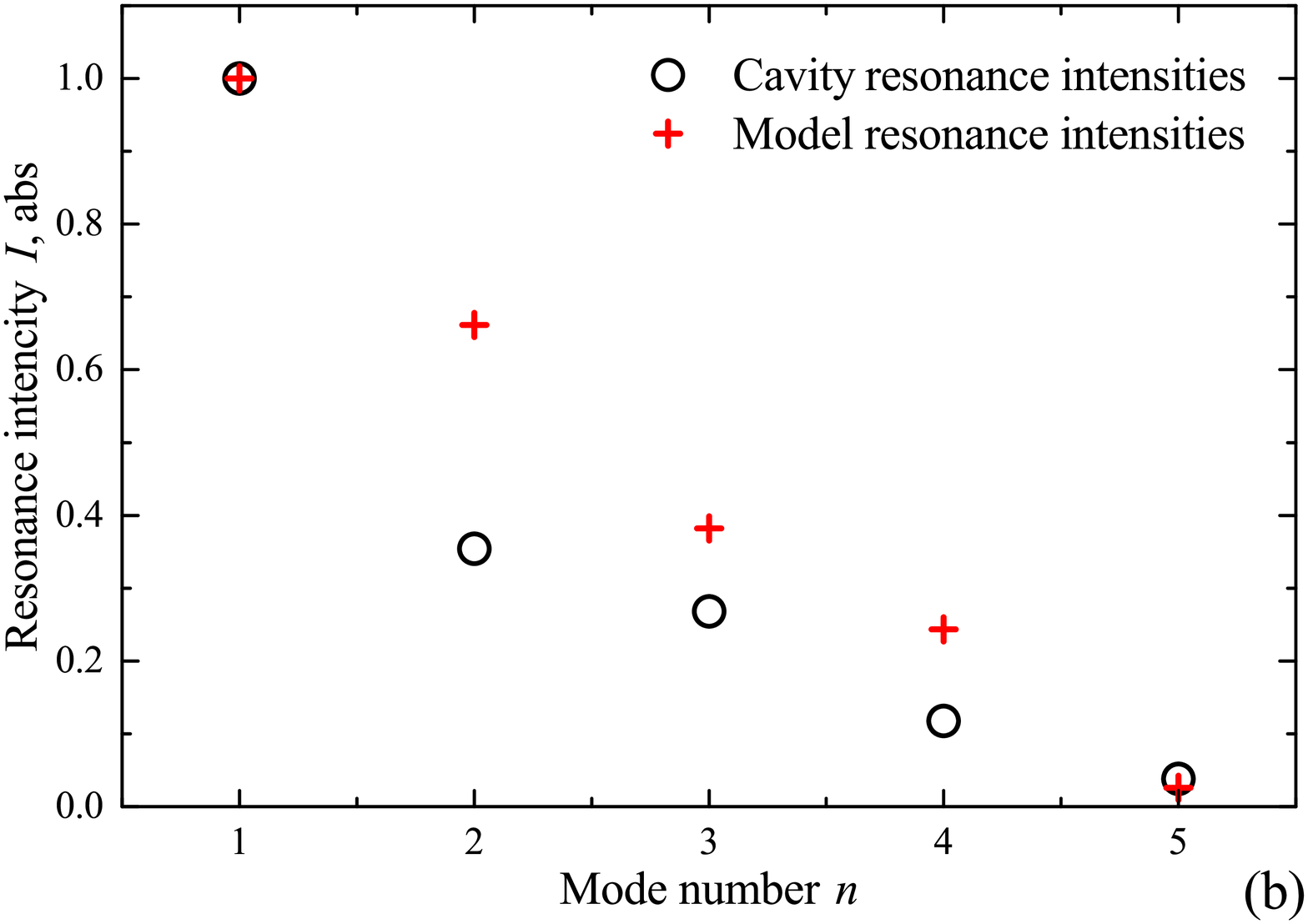}
\includegraphics[width=0.67\columnwidth]{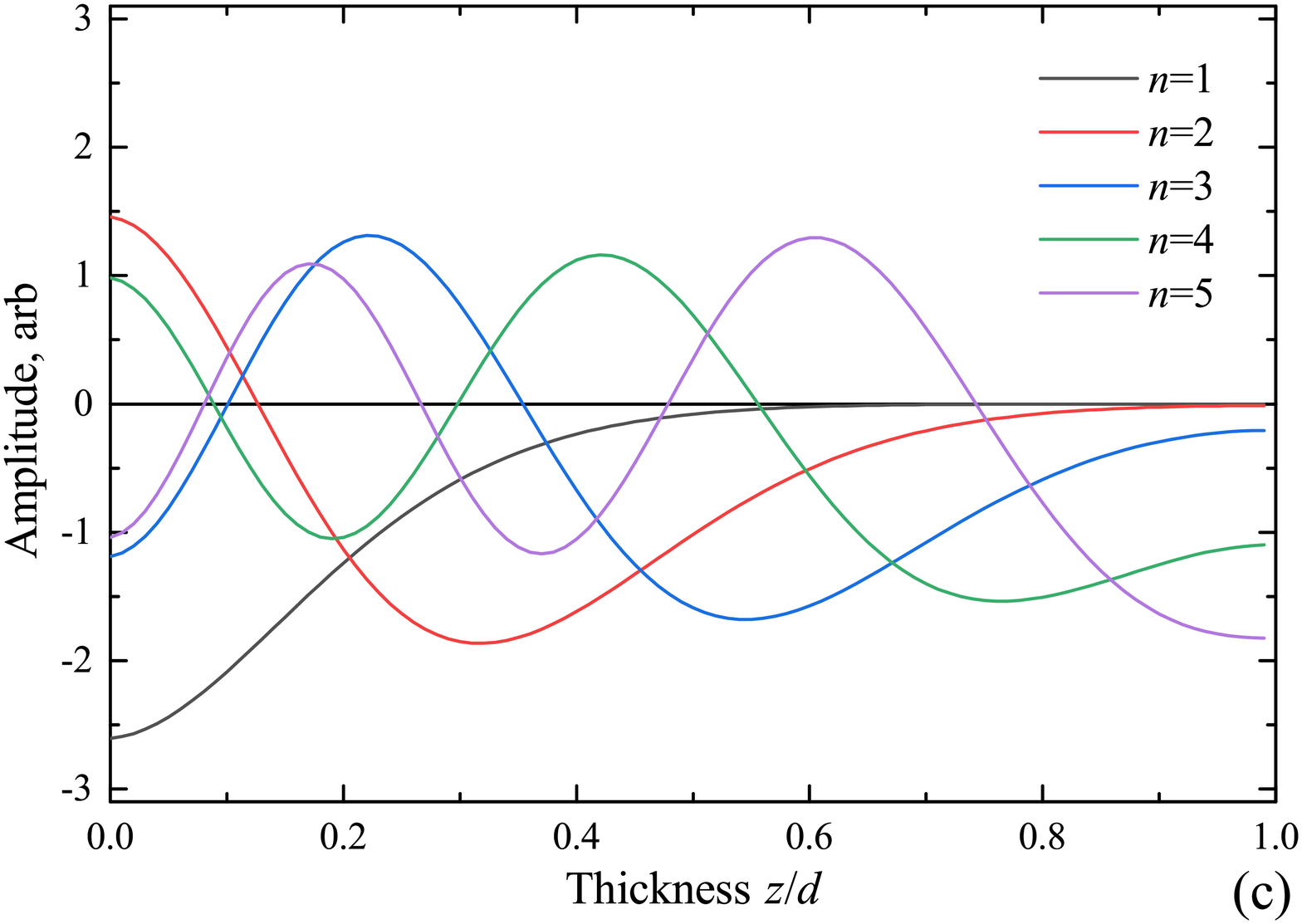}
\caption{a) Dependencies of resonance fields on the mode number $\mu_0H_r(n)$ at $f=9.44$~GHz.
b) Dependencies of resonance intensities on the mode numbers $I(n)$.
c) Dependencies of amplitudes of magnetization precession across the thickness of the film for different modes $n$. }
\label{Theory}
\end{center}
\end{figure*}

Resonance modes $n=1,.,5$ in Figs.~\ref{FMR_exp},~\ref{EPR_exp},~and~\ref{FMR_sum} are refereed to as the perpendicular standing spin wave (PSSW) resonances \cite{Kittel_PR_100_1295,Seavey_JAP_30_S227,Klingler_JPDAP_48_015001} where non-zero spin wavevector induces effective exchange field.
Figure~\ref{Theory}a shows the dependence of the CMR resonance field on the mode number $H_r(n)$, 
Figure~\ref{Theory}b shows the dependence of the relative CMR amplitude on the mode number $I(n)$.
For conventional exchange PSSW the dependence $H_r(n)$ is parabolic due to $k^2$ factor of the $H_{ex}$ in Eq.~\ref{kit}.
In contrast, the dependence $H_r(n)$ in Fig.~\ref{Theory}a is approximately linear, which indicates that spin wave resonances can be characterised as anomalous \cite{Portis_APL_2_69,Sasaki_JSNM_16_14,Rappoport_PRB_69_1252133}.
Anomalous spin wave resonances appear when any magnetic property, i.e., the saturation magnetization, magnetic anisotropies, or the exchange stiffness, is non-uniform across the thickness of ferromagnetic film.
In our case all three parameters are non-uniform due to the gradient composition of the Pd-Fe film.

PSSW in ferromagnetic films can be treated comprehensively by means of the collective dynamic equation \cite{Portis_APL_2_69,Sasaki_JSNM_16_14,Rappoport_PRB_69_1252133,Puszkarski_ProgSurfSci_9_191}.
It can be shown that by starting from the Landau–Lifshitz equation for an individual layer inside the film coupled to neighbouring layers via the exchange interaction the dynamic equation is derived in the form of the Schrodinger equation
\begin{equation}
\left[-D(z)\frac{\partial^2}{\partial z^2}+V(z)\right] m(z)=\frac{2\pi f_r}{\gamma} m(z), 
\label{S}
\end{equation}
where $m(z)$ is the dependence of the locally-normalized amplitude of magnetization precession across the thickness of the film, $D(z)=2A(z)/\mu_0M_s(z)$ is the exchange coefficient, and $V(z)=H-M_{eff}(z)+D(z)/M_s(z)\frac{\partial^2}{\partial z^2}M_s(z)$ is the potential well.
The general, boundary conditions of the eigenvalue problem in Eq.~\ref{S} are \cite{Soohoo_PRB_131_594,Bajorek_JAP_42_4324,Bailey_PRB_8_3247,Puszkarski_ProgSurfSci_9_191}
\begin{equation}
dm/dz+\alpha_s m=0, 
\label{BC}
\end{equation}
where $\alpha_s=K_s/A_s$ is the surface coefficient, $K_s$ is the surface anisotropy, $A_s$ is the exchange stiffness constant at the surface.
In case if $\alpha_s=0$ the spin boundary conditions are opened. 
In case if $1/\alpha_s=0$ the spin boundary conditions are closed or pinned.
The intensity of the resonance is $\propto \int{m(z)M_s(z)dz}$. 

CMR resonances are analysed using Eqs.~\ref{S} and \ref{BC} as follows. 
We consider the linear dependence of concentration of Fe $C_{\rm Fe}$ across the thickness: $C_{\rm Fe}=10.8-7.5z/d$ at.\%.
The local saturation magnetization $M_s(z)$ is found as $C_{\rm Fe}\times\overline{M}_s/7.1$.
Next, it is assumed that the local effective magnetization is proportional to the local saturation magnetization $M_{eff}(z)=\beta_K M_s(z)$, where $\beta_K<1$ corresponds to the easy out-of-plane anisotropy and vice versa.
At last, it can be noted that due to the linear dependence of both the saturation magnetization and the Curie temperature on Fe composition \cite{Ewerlin_JPCM_25_266001,Esmaeili_SciChiMat_64_1246} the exchange coefficient $D$ can be considered as independent of composition and, thus, constant across the thickness of the film.

In numerical calculations we optimise resonance fields of spin wave resonances (see Fig.~\ref{Theory}a).
The optimum fit is obtained with $D=15$~T$\cdot$nm$^2$, $\beta_K=1.16$, and open boundary conditions. 
The exchange coefficient is well consistent with with typical values for ferromagnetic metals.
For instance, in permalloy with $A=1.2\times10^{-11}$~J/m and $\mu_0M_s=1$~T the exchange coefficient is $D=30$~T$\cdot$nm$^2$.
The obtained $\beta_K=1.16>1$ indicates that the hard axis is aligned out-of-plane, which is well consistent with previous studies of epitaxial Pd-Fe films \cite{Esmaeili_SciChiMat_64_1246}.
The obtained open boundary conditions are in agreement with our FMR studies of Pd-Fe films with uniform composition \cite{Esmaeili_SciChiMat_64_1246}: for such films we have not observed PSSW resonances.
Figure~\ref{Theory}b shows the dependence of resonance amplitudes on the mode number. 
Both experimental and numerical $I(n)$ decrease progressively with the mode number and show comparable values. 
Figure~\ref{Theory}c shows dependencies of the locally-normalized amplitude of magnetization precession on position $m(z)$ for first four modes.


%
\begin{figure*}[!ht]
\begin{center}
\includegraphics[width=0.67\columnwidth]{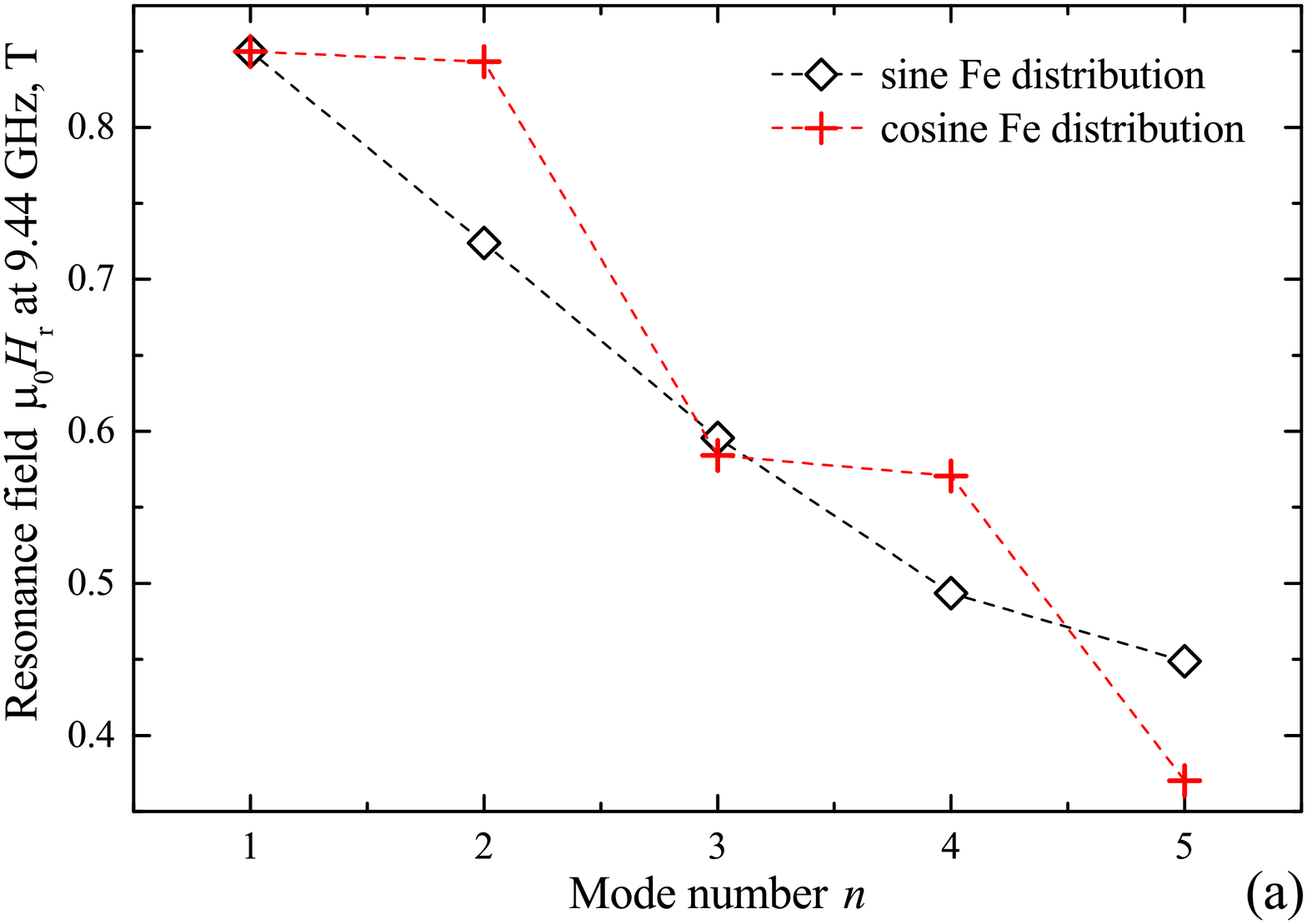}
\includegraphics[width=0.67\columnwidth]{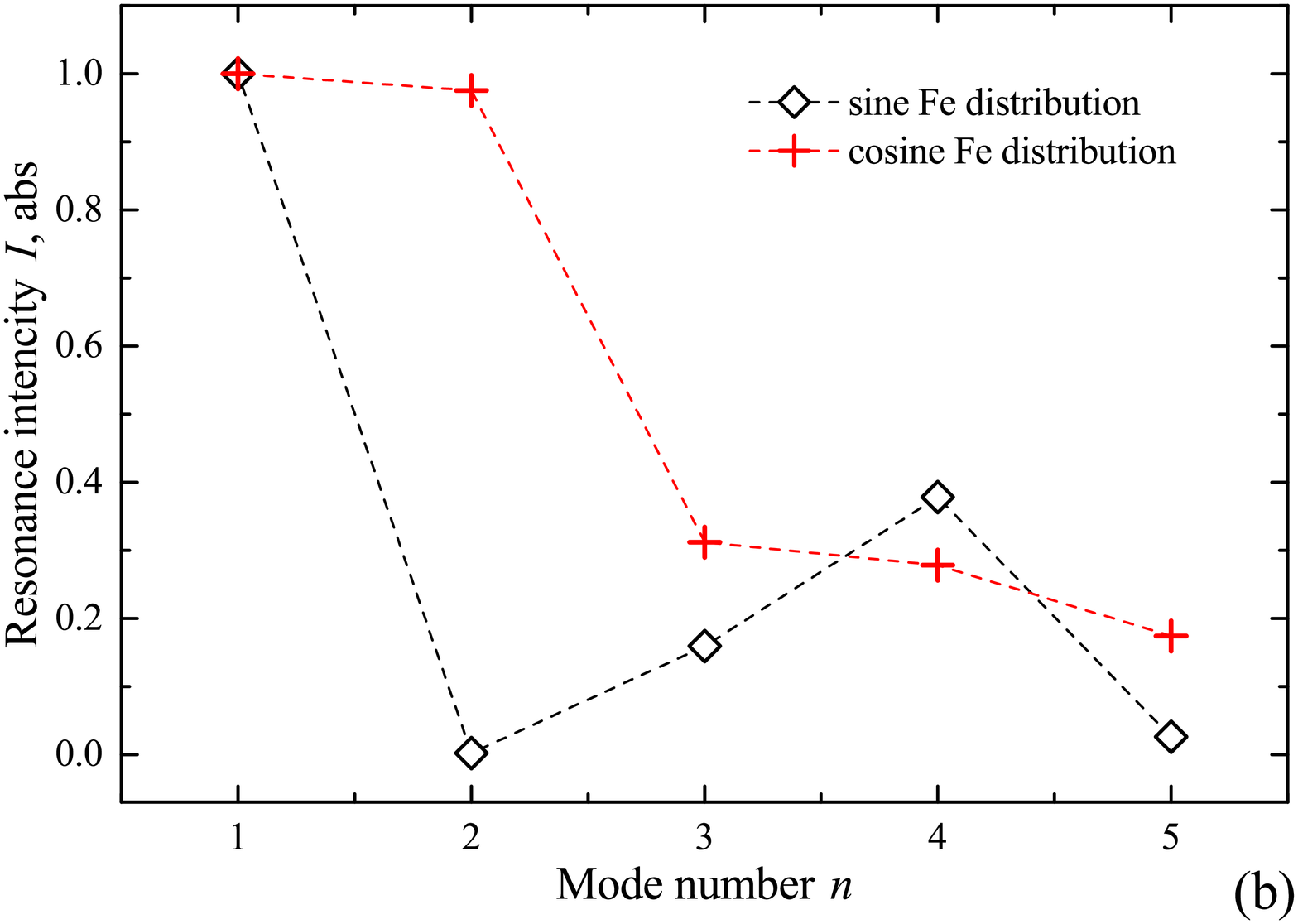}
\includegraphics[width=0.67\columnwidth]{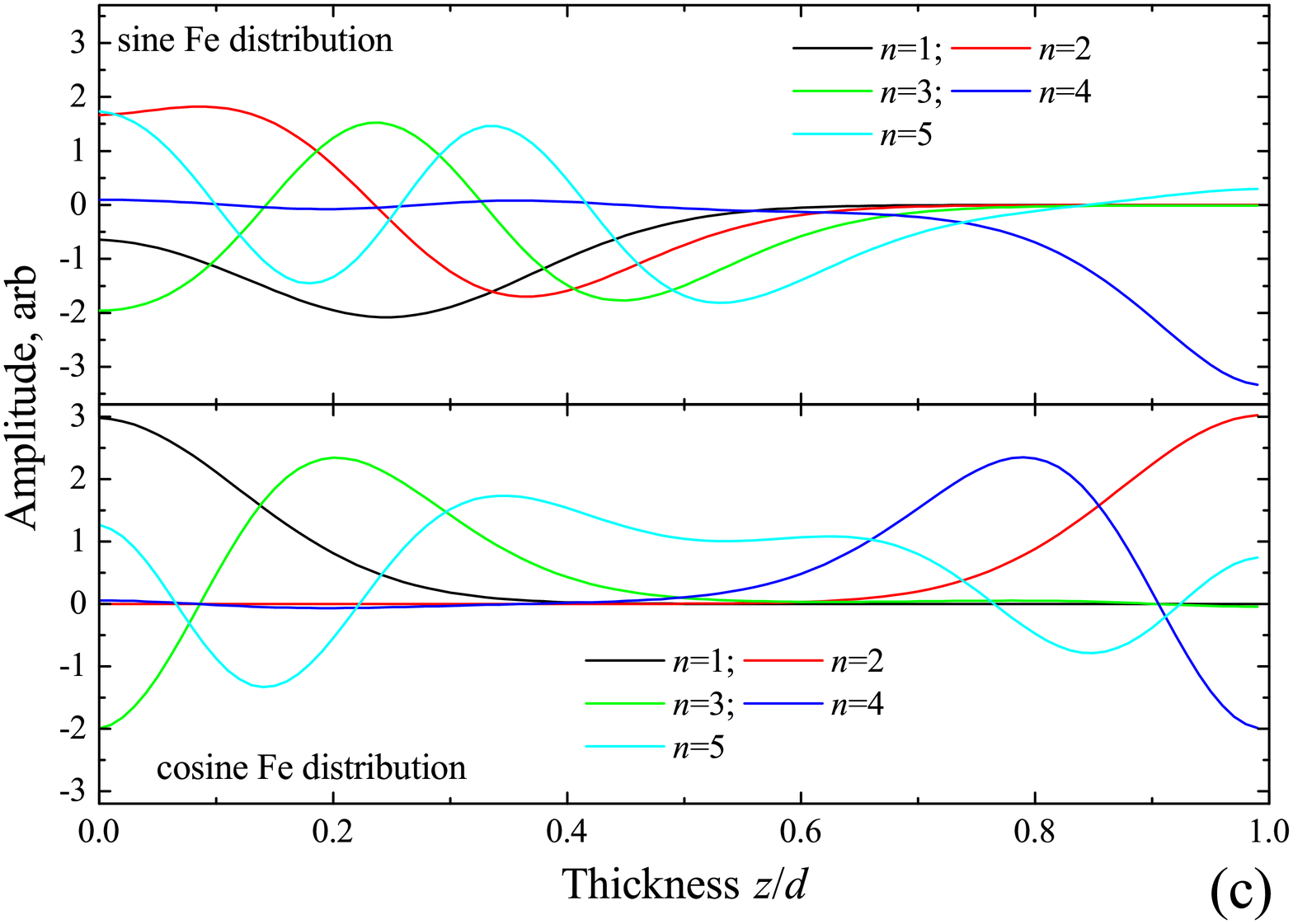}
\caption{a) Dependencies of resonance fields on the mode number $\mu_0H_r(n)$ at $f=9.44$~GHz.
b) Dependencies of resonance intensities on the mode numbers $I(n)$.
c) Dependencies of amplitudes of magnetization precession across the thickness of the film for different modes $n$.}
\label{Theory_ex}
\end{center}
\end{figure*}

For demonstration purposes we consider sine and cosine distributions of Fe across the thickness of 100 nm film: $C_{\rm Fe}=10+8\sin(2\pi z/d)$ at.\%, and $C_{\rm Fe}=10+8\cos(2\pi z/d)$ at.\%, with the same parameters $D$ and $\beta_K$ as in Fig.~\ref{Theory}.
For considered compositions the maximum, the minimum and the average Fe content are identical.
Yet, resonance contritions for PSSW in these films are different.

Figure~\ref{Theory_ex} shows resonance fields, intensities and distributions of amplitudes of magnetization precession for considered compositions.
The sine composition shows monotonous decrease of resonance field with the mode number, while for cosine composition this dependence is step-wise.
For sine composition the strongest modes are $n=1$ and $n=4$, for cosine composition the strongest modes are $n=1$ and $n=2$.
These differences highlight the potential of engineering of ferromagnetic thin films with nonuniform composition for high-frequency or low-field magnonic operations.
%

\section{Conclusion}

Summarising, we have studied ferromagnetic resonance phenomenon in ferromagnetic thin films with essentially non-uniform composition.
The Pd-Fe thin film with linear distribution of Fe content across the thickness was used as the model material.
Anomalous perpendicular standing spin waves were observed using cavity magnetic resonance spectroscopy and VNA-FMR spectroscopy.  
Numerical analysis of resonance conditions using the collective dynamic equation yields the exchange stiffness constant for diluted Pd-Fe alloy $D=2A/\mu_0M_s=15$~T$\cdot$nm$^2$ and the ratio of the effective magnetization to the saturation magnetization $M_{eff}/M_s=1.16$.
As a demonstration, perpendicular standing spin waves have been considered numerically in Pd-Fe films with sine and cosine distributions of Fe content. 
Overall, engineering of thin films with non-uniform composition across the thickness can be used for high-frequency or low-field magnonic operations using exchange spin waves.

\section{Acknowledgements}

This work was supported by the Russian Science Foundation (Project N 22-22-00629).

\bibliography{A_Bib_FePt_grad}

\end{document}